\documentclass[a4paper]{article}

\usepackage[utf8]{inputenc}
\usepackage[english]{babel}

\usepackage[cm]{fullpage}
\usepackage{amsfonts}
\usepackage{array}
\usepackage{graphicx}
\usepackage{wrapfig}
\usepackage{xcolor}
\graphicspath{{images/}}
\usepackage{multicol}
\usepackage{multirow}
\usepackage{ragged2e}
\usepackage{float}
\usepackage{hyperref}
\usepackage[all]{hypcap}
\usepackage{placeins}
\usepackage{subcaption}
\usepackage[
  backend=biber,
  style=nature,
  sorting=none
]{biblatex}
\usepackage{bm}
\usepackage{caption}
\usepackage{fancyhdr}
\usepackage{authblk}
\usepackage{titling}
\newenvironment{Figure}
  {\par\medskip\noindent\minipage{\linewidth}}
  {\endminipage\par\medskip}

\title{\textbf{Configuration-dependent electronic and optical properties of 2D Mo$_{1-x}$W$_x$S$_2$ alloys across the full composition range}}

\author{
  M. Szyszko $^{1}$, M. Birowska $^{1}$ \\
  \normalsize{$^{1}$University of Warsaw, Faculty of Physics, 00-092 Warsaw, Pasteura 5, Poland}
}

\date{} 
\begin{document}
\begin{refsection}

\maketitle

\hrule
\section*{\normalsize ABSTRACT}
Here we analyze multiple symmetry-inequivalent atomic configurations across the entire composition range of the isovalent and isostructural Mo$_x$W$_{1-x}$S$_2$ alloy using density-functional theory and Monte Carlo simulations. Our results show that although structural stability and energetics are largely composition-driven, the electronic and optical properties exhibit  configuration dependence, with local atomic arrangements critically shaping band-edge splitting, valley structure, effective-mass anisotropy, and optical selection rules. In contrast to the pristine monolayers, even in the absence of spin-orbit coupling (SOC), splitting of the band edges at the $K$ point is observed across the entire composition range. In particular, while the valence-band maximum (VBM) remains largely robust, the conduction-band minimum (CBM) shows strong configuration-dependent splitting from few meV up to hundredths of meV. This behaviour leads to a non-trivial dependence of the valley energetics. Configurations with well-separated conduction bands support additional optically active transitions beyond the conventional A and B excitons in MoS$_2$ and WS$_2$ monolayers, whereas nearly degenerate cases exhibit a reduced number of allowed transitions, observed for specific configurations at $x = 1/3$ and $x = 2/3$. These results demonstrate that the number and character of optically active transitions are governed not only by composition, but also by the microscopic arrangement of atoms. Moreover, we found the hole effective masses at the VBM show  configuration-dependent anisotropy, reflecting sensitivity to local symmetry breaking and implying direction-dependent transport.

\hrule
\begin{multicols}{2}

\section{\normalsize INTRODUCTION}
Two-dimensional (2D) transition-metal dichalcogenides (TMDs) such as MoS$_2$ and WS$_2$ have emerged as versatile platforms for next-generation nanoelectronics, optoelectronics, valleytronics, and catalysis~\cite{radisavljevic2011single, wang2012electronics, xiao2012coupled, mak2012control, xu2022frenkel, huang2019atomically}. Their direct band gaps in the visible range~\cite{mak2010atomically}, strong spin--orbit coupling (SOC)~\cite{zhu2011giant}, and pronounced many-body effects~\cite{chernikov2014exciton} enable efficient light--matter interaction and control of charge, spin, and valley degrees of freedom within a single atomic layer. However, the discrete band-edge energies and fixed lattice parameters of the pristine compounds limit continuous tunability of their properties.

Isostructural and isovalent alloying in Mo$_{1-x}$W$_x$S$_2$ provides a powerful route to overcome these limitations by enabling continuous tuning of structural, electronic, and optical properties while preserving crystal symmetry and chemical compatibility~\cite{dumcenco2013visualization}. While such alloys are often described in terms of composition-dependent trends, the role of atomic-scale dopant configurations remains far less explored. In particular, key quantities such as band-edge energies, effective masses, valley energetics, optical selection rules, and excitonic properties can exhibit significant sensitivity to the local atomic arrangement, which is not captured by averaged or virtual-crystal approaches.

Previous experimental and theoretical studies have primarily focused on selected compositions or have neglected the combined effects of SOC and configurational variability~\cite{tan2017ordered, xia2021atomic}. As a result, a systematic understanding of how both composition and atomic-scale disorder influence the electronic structure and optical response across the full alloy range is still lacking. This is particularly important for technologically relevant properties, such as the number of optically active transitions, band-edge splittings, and valley-dependent optical selection rules.

Here, we address this gap by performing a comprehensive first-principles study of monolayer Mo$_{1-x}$W$_x$S$_2$ across the entire composition range. Using a $3 \times 3$ supercell, we consider all symmetry-inequivalent configurations for ten distinct W concentrations, resulting in a consistent data set that allows us to disentangle composition-driven trends from configuration-induced variations. This approach enables us to quantify the extent to which local atomic arrangements affect band dispersion, spin--orbit splittings, effective masses, valley energetics, and the number of optically active transitions.

In addition, the resulting first-principles database provides a foundation for a cluster-expansion Hamiltonian describing the energetics of the alloy. Monte Carlo simulations based on this Hamiltonian allow us to extend the analysis to experimentally relevant length and time scales, capturing temperature-dependent short-range order and its impact on macroscopic observables. Taken together, our multiscale approach provides a unified picture of how composition and atomic configuration jointly determine the energetics, structural, electronic and optical properties of Mo$_{1-x}$W$_x$S$_2$.

\end{multicols}
\begin{Figure}
    \centering
    \includegraphics[scale=0.09]{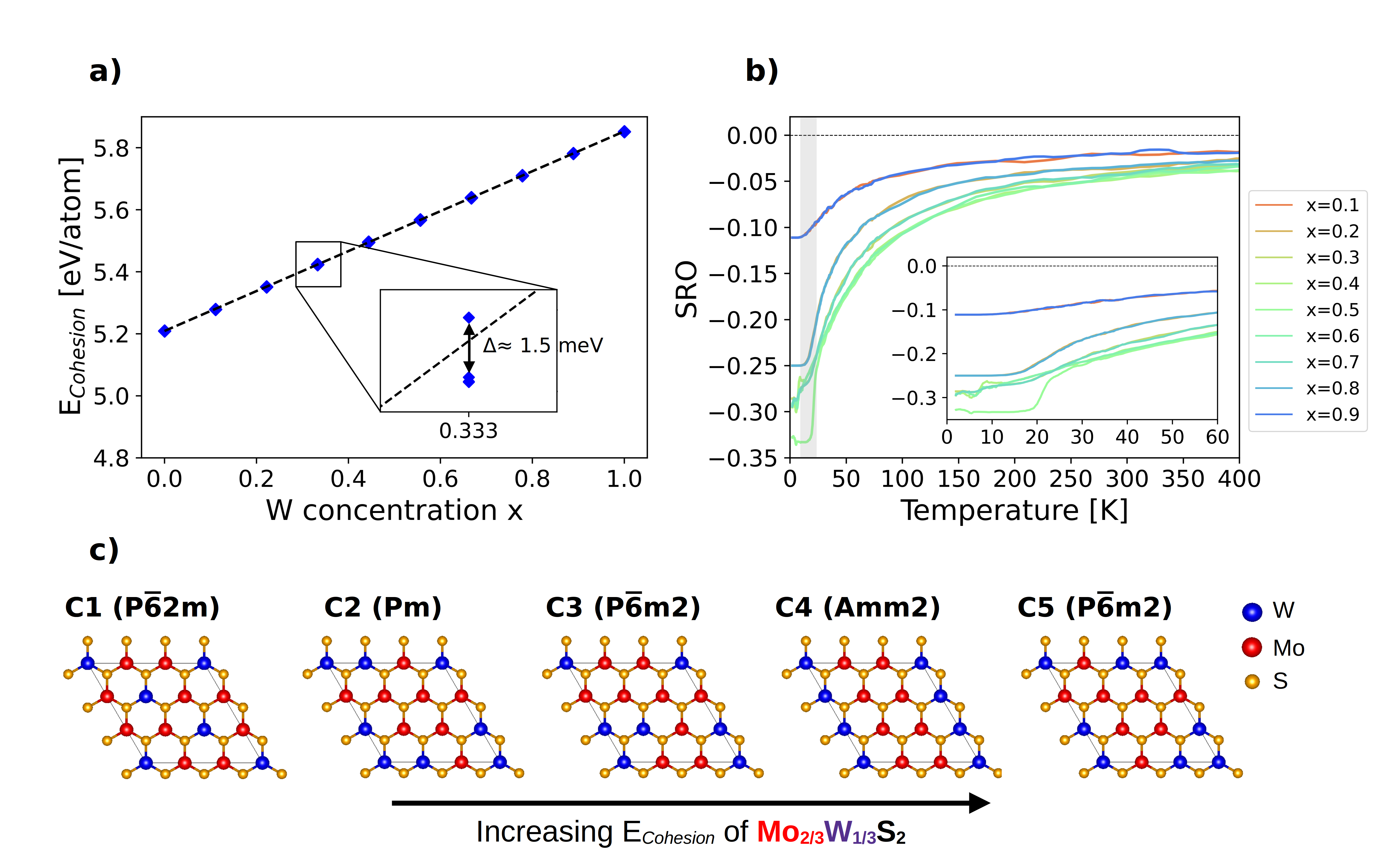}
    \captionof{figure}{Energetics of the Mo$_{1-x}$W$_x$S$_2$ alloy.
(a) Cohesive energy as a function of W concentration \(x\), showing an approximately linear dependence. Minor deviations between different atomic configurations at fixed \(x\) are on the order of \(\sim 1.5\) meV.
(b) Binary short-range order (SRO) parameter obtained from Monte Carlo simulations performed on \(20\times20\) supercells, exhibiting similar behaviour for all concentrations \(x\): an ordered state at low temperatures and an order--disorder transition occurring around \(T \approx 20\) K.
(c) All symmetry-inequivalent configurations for $x=1/3$, demonstrating negligible dependence of the total energy on global symmetry and a weak energetic preference for dopant separation, with configurations exhibiting enhanced dopant clustering being slightly higher in energy. The C3 and C5 configurations exhibit the same space group symmetry as pristine MoS$_2$ and WS$_2$.
}
    \label{Energetics}
\end{Figure}

\begin{multicols}{2}
\section{\normalsize RESULTS AND DISCUSSION}
In this study, we systematically examine the impact of dopant concentration and spatial distribution in the 2H phase of Mo$_{(1-x)}$W$_{x}$S$_{2}$ alloy on its energetic, electronic, and optical properties. 

\subsection{Structural properties and energetics.}

We begin by analysing the cohesive energies of the investigated systems, shown in Fig.~\ref{Energetics}. For each of the 28 configurations considered, all ionic positions were fully relaxed, while the lattice parameters for a given concentration \(x\) were interpolated between those of the pure MoS\(_2\) and WS\(_2\) compounds. Owing to the small lattice-parameter mismatch between MoS\(_2\) and WS\(_2\) (approximately 0.2 \%), the residual stress tensor in all calculations was negligible.

As shown in Fig.~\ref{Energetics}(a), the cohesive energy exhibits a clear linear dependence on composition. This behaviour indicates that Mo\(_{1-x}\)W\(_x\)S\(_2\) behaves approximately as an ideal or weakly interacting solid solution, for which the cohesive energy can be well approximated by a linear interpolation between the end members,
\begin{equation}
E_{\mathrm{coh}}(x) \approx (1 - x)\, E_{\mathrm{coh}}^{\mathrm{MoS_2}} + x\, E_{\mathrm{coh}}^{\mathrm{WS_2}} .
\end{equation}
The near-linearity of \(E_{\mathrm{coh}}(x)\) demonstrates that the energetics are dominated by composition, with only minor configuration-dependent corrections. For a fixed concentration, the energy differences between distinct atomic configurations do not exceed 2~meV (see Fig. \ref{Energetics}), which is about two orders of magnitude smaller than the energy differences between different concentrations.

While composition controls the dominant energetic trends, analysing the correlation between total energy and atomic configuration provides additional insight into the thermodynamic behaviour of the alloy. We find that the total energy is largely insensitive to the global symmetry of the dopant arrangement (C3 configuration exhibits the  same space group symmetry as C5, as presented in Fig. \ref{Energetics}(c)); instead, it is primarily governed by local chemical environments. In particular, configurations in which dopant atoms form nearest-neighbour pairs are slightly energetically unfavourable. This reflects a weak but finite energetic penalty associated with like-atom nearest-neighbour interactions.

At intermediate to high dopant concentrations, \(x \in (1/3,\,2/3)\), complete avoidance of dopant--dopant nearest neighbours is not possible within the finite \(3\times3\) supercell used in the DFT calculations, such that some degree of local clustering is unavoidable. In this concentration regime, previous studies \cite{tan2017ordered} have suggested the emergence of an order--disorder transition, driven by competition between the small energetic preference for homogeneous mixing and the rapidly increasing configurational entropy.

To explicitly account for thermal disorder and extend the analysis beyond ordered configurations, we constructed a cluster expansion model parametrised using 215 additional DFT calculations performed for \(4\times4\) supercells. This effective Hamiltonian enables efficient sampling of the configurational phase space at finite temperatures. The thermodynamic behaviour was subsequently investigated using Monte Carlo simulations on a \(20\times20\) supercell over a temperature range from 1~K to 400~K. The details of the cluster-expansion procedure have been described in Section 3 of Supporting Information (SI).

The order--disorder transition was characterised by evaluating the short-range order (SRO) parameter at each Monte Carlo step, as shown in Fig.~\ref{Energetics}(b). The SRO was defined using first-neighbour correlations and evaluated with respect to Mo atoms, with negative values indicating a tendency towards homogeneous mixing and positive values signalling a tendency towards phase separation. An equivalent behaviour is obtained when defining the SRO with respect to W atoms. At low temperatures, the system remains ordered for all concentrations, exhibiting a preference for homogeneously distributed molybdenum atoms, as evidenced by the negative SRO values. With increasing temperature, an order--disorder transition occurs in the range of approximately 15--25~K. In the high-temperature limit, the SRO parameter asymptotically approaches zero, corresponding to a fully random distribution of transition-metal atoms, which matches previously reported results \cite{tan2017ordered, gan2014order}.

These results imply that at typical TMD growth temperatures (in the range of 400-1000 K at various stages of synthesis) \cite{zhao20212d, maurtua2024molecular}, Mo\(_{1-x}\)W\(_x\)S\(_2\) alloys are expected to form in a thermodynamically disordered phase. This conclusion is in good agreement with experimental observations reporting random or only weakly correlated distributions of Mo and W atoms in grown samples~\cite{dumcenco2013visualization}.

\end{multicols}
\begin{Figure}
    \centering
    \includegraphics[scale=0.09]{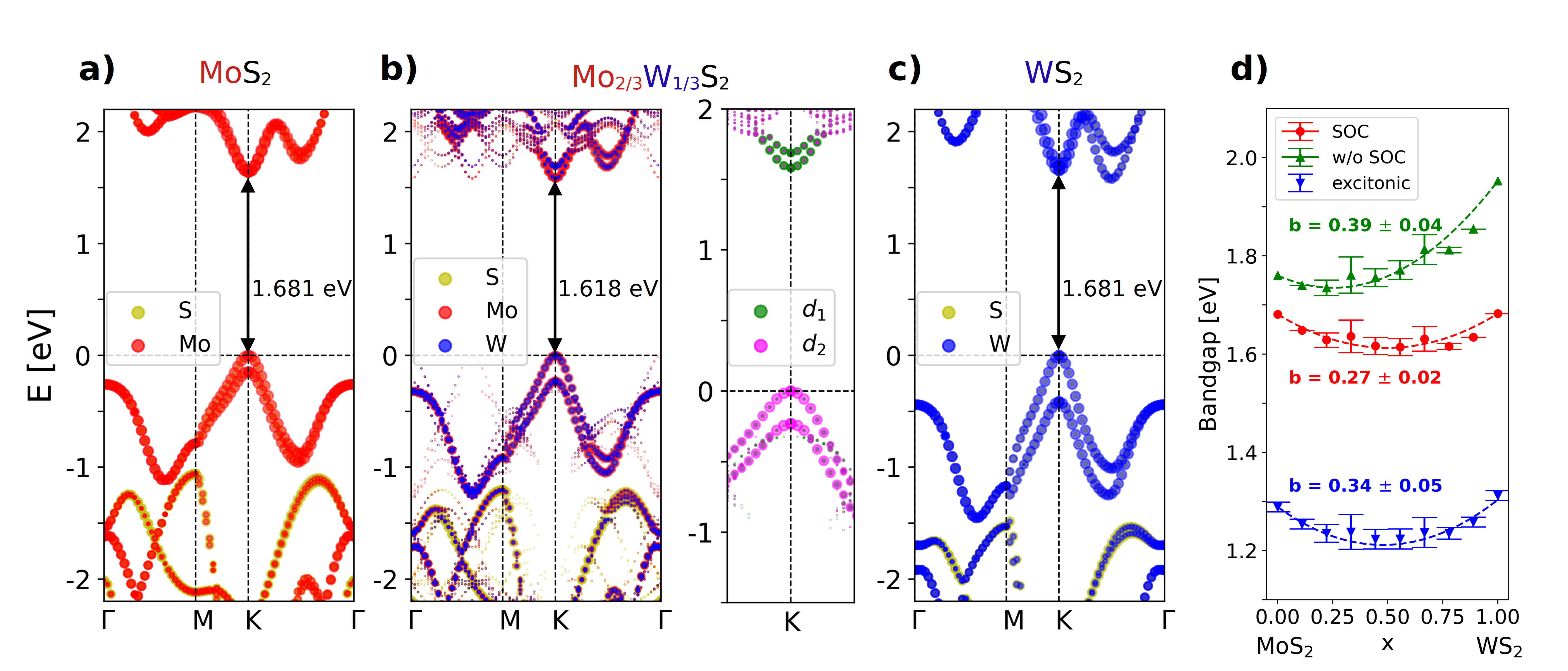}
    \captionof{figure}{PBE+SOC band structures of (a) $MoS_{2}$, (b) the C2 configuration of $Mo_{2/3}W_{1/3}S_{2}$, and (c) $WS_{2}$, with atomic projections (red: Mo, blue: W, yellow: S). The alloy preserves the direct gap at $K$. In all systems, both VBM and CBM are dominated by transition-metal states: the VBM mainly by $d_{x^{2}-y^{2}} + d_{xy}$ ($d_{2}$) orbitals, and the CBM by $d_{z^{2}}$ ($d_{1}$) orbitals. The VBM reflects the alloy composition, while the CBM depends sensitively on composition and atomic arrangement. For WS$_2$, we note that the $Q$ valley lies lower in energy than the $K$ valley. This behavior is likely related to the chosen lattice constant of 3.153~\AA\ and the inclusion of SOC, whereas PBE calculations without SOC correctly predict the CBM at the $K$ point. (d) Bandgap versus composition: points denote the mean bandgap over configurations, with error bars indicating standard deviation of the bandgaps. The pronounced nonlinearity is quantified by the bowing parameter $b$ [in eV] for  PBE (green), PBE+SOC (red) and PBE+SOC (blue) including exciton binding energy (Rytova–Keldysh model, procedure and the results described in Section 4 of SI).}
    \label{Band structures}
\end{Figure}
\begin{multicols}{2}   

\subsection{\normalsize ELECTRONIC AND OPTICAL PROPERTIES}

\subsubsection*{Band gaps}
The electronic band structures of the alloys were calculated using $3 \times 3$ supercells and subsequently unfolded onto the primitive cell for a set of ten compositions $x$. The band structure corresponding to $x = 1/3$ is shown in Fig.~\ref{Band structures}(b). For all considered systems, the fundamental band gap remains direct at the $K$ point (with the exception for WS$_{2}$), with its value varying systematically with W concentration $x$.  For a given composition, noticeable configuration-dependent variations of the direct band gap are observed, reaching up to $\sim 100$~meV for $x = 1/3$ and $x = 2/3$, as illustrated by the error bars in Fig.~\ref{Band structures}(d). These variations correlate with the total energy: the lowest-energy configurations consistently exhibit the smallest band gaps, while higher-energy configurations show larger values. This trend is reflected in Fig.~\ref{Band structures}(d), where the lower bounds of the error bars correspond to the lowest-energy configurations.

The gap exhibits a clear non-linear behaviour that can be described by the standard bowing relation \cite{mourad2012theory}
\begin{equation}
E_g(x) = (1-x)E_g^{\mathrm{MoS_2}} + xE_g^{\mathrm{WS_2}} - x(1-x)b,
\end{equation}
where \(b\) is the bowing parameter.
The obtained values are relatively small, amounting to 0.39~eV within PBE and decreasing to 0.27~eV upon inclusion of spin--orbit coupling, the latter being in better agreement with available theoretical and experimental data \cite{tan2017ordered, wang2016chemical, chen2013tunable}. After accounting for the exciton binding energy (see Fig. S2), the bowing of the optical gap increases to 0.34~eV, indicating an additional, composition-dependent contribution arising from excitonic effects. This highlights the importance of distinguishing between electronic and optical bowing in low-dimensional systems, where excitonic effects are pronounced. 
In comparison to conventional semiconductor alloys such as InGaN or GaAsN, where bowing parameters typically range from $\sim$1~eV to several eV \cite{mourad2010band, wei1996giant}, the present values are significantly smaller. Instead, they are comparable to weakly mismatched systems such as AlGaAs and GaAsP, which exhibit bowing below 1~eV \cite{vurgaftman2001band}. This places the studied system in the small-bowing regime, characteristic of chemically similar and structurally compatible alloys. Such behavior enables predictable band-gap engineering with reduced sensitivity to disorder, which is advantageous for heterostructure design and optoelectronic applications \cite{alferov2001nobel}.

\subsubsection*{Effective masses}
Furthermore, a nearly linear dependence of the carrier effective masses on composition is observed. In general, hole effective masses are slightly larger than those of electrons. For the end compounds, the calculated values are in good agreement with previous theoretical and experimental reports, yielding electron (hole) effective masses of $\sim$0.4--0.6~$m_e$ ($\sim$0.5--0.7~$m_e$) for MoS$_2$ and $\sim$0.3--0.4~$m_e$ ($\sim$0.35--0.5~$m_e$) for WS$_2$~\cite{tanabe2016band, yu2015phase, tan2017ordered}. Compared to conventional semiconductors such as GaAs or GaN~\cite{vurgaftman2001band}, these values are relatively large, reflecting the flatter band dispersion of $d$-orbital-derived states. The similar magnitudes of electron and hole effective masses suggest comparable carrier mobilities, which is favourable for balanced charge transport and efficient recombination.

\begin{Figure}
    \centering
    \includegraphics[scale=0.16]{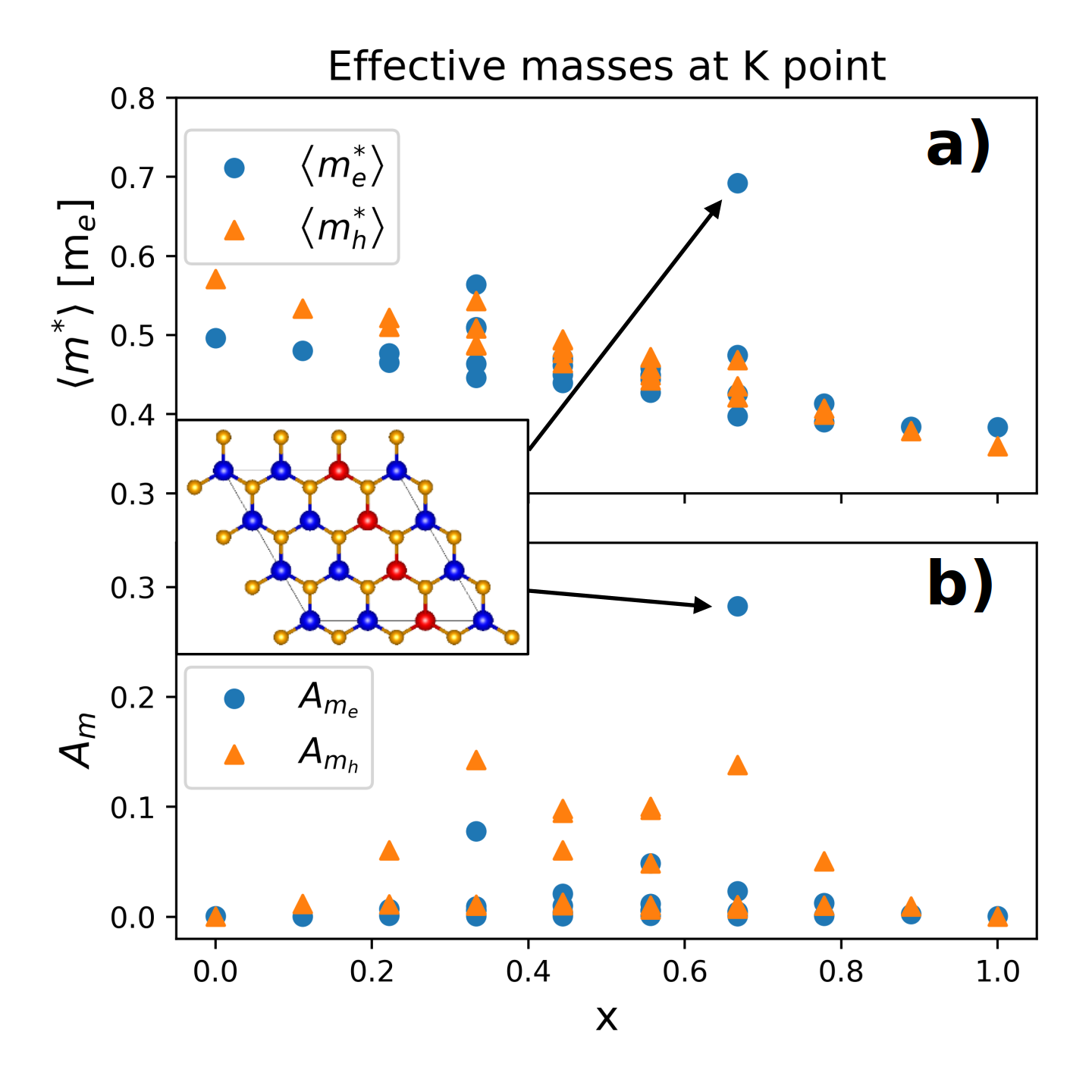}
    \captionof{figure}{The in-plane effective masses of charge carriers calculated at the $K$ high-symmetry point (top panel), together with band-edge effective mass anisotropy $A_{m}$ (bottom panel). The electron and hole effective masses are evaluated at the conduction-band minimum (CBM) and valence-band maximum (VBM), respectively. The effective masses exhibit an approximately linear dependence for both types of carriers. Deviations from this trend arise from structural anisotropy associated with specific atomic configurations as pointed by the inset.}
    \label{effective}
\end{Figure}

For a given composition, the effective masses show a noticeable dependence on the atomic configuration, with variations up to $\sim$20\% across non-equivalent structures. To quantify directional effects, we define the band-edge effective-mass anisotropy parameter as
\begin{equation}
\mathrm{A_{m}} = \frac{m_{\max} - m_{\min}}{m_{\max}},
\end{equation}
where $m_{\max}$ and $m_{\min}$ are the in-plane effective masses along the principal axes in reciprocal space. The $A_{m}$ remains small for most compositions, particularly for the conduction band, indicating nearly isotropic transport. In contrast, the valence band exhibits enhanced anisotropy in selected configurations, which can be attributed to local symmetry breaking induced by specific dopant arrangements.

The presence of effective-mass anisotropy implies direction-dependent carrier transport, with higher mobility along directions of lower effective mass, leading to anisotropic conductivity.

\end{multicols}

\begin{table}[ht]
\centering
\caption{Structural and electronic properties of symmetry-inequivalent configurations for the Mo$_{2/3}$W$_{1/3}$S$_2$ alloy. $\Delta E_{\mathrm{VBM}}^{\mathrm{SP}} / \Delta E_{\mathrm{VBM}}^{\mathrm{SOC}}$ (meV) and $\Delta E_{\mathrm{CBM}}^{\mathrm{SP}} / \Delta E_{\mathrm{CBM}}^{\mathrm{SOC}}$ (meV) denote the energy splitting of the valence-band maximum (VBM) and conduction-band minimum (CBM) at the $K$ point, obtained without and with spin--orbit coupling, respectively. The atomic ($w^{\mathrm{VBM}}$, $w^{\mathrm{CBM}}$) and orbital characters refer to the dominant contributions to the VBM and CBM states. The quantity $\Delta^{\mathrm{VBM}}_{K\Gamma} = E^{\mathrm{VBM}}(K) - E^{\mathrm{VBM}}(\Gamma)$ and $\Delta^{\mathrm{CBM}}_{KQ} = E^{\mathrm{CBM}}(Q) - E^{\mathrm{CBM}}(K)$ give the energy difference between the VBM at the $K$ and $\Gamma$ points and CBM at $Q$ and $K$ points, respectively. The number of active transitions corresponds to band-edge transitions between VBM and CBM states with significant dipole matrix elements (i.e., nonzero oscillator strength). The dipole selection rules for circularly polarized light are given by $\left| \hat{\mathbf{e}}_{\pm} \cdot \mathbf{p}_{cv\mathbf{k}} \right|^2 > 0 \leftrightarrow \sigma^{\pm}$, where $\hat{\mathbf{e}}_{\pm} = \frac{1}{\sqrt{2}} \left( \hat{\mathbf{x}} \pm i \hat{\mathbf{y}} \right)$ correspond to right- ($\sigma^{+}$) and left- ($\sigma^{-}$) circular polarization, respectively. Pristine monolayers MoS$_2$ and WS$_2$ exhibit P$\overline{6}$m2 (187) space group symmetry. The last row reports the band-edge effective-mass anisotropy parameter, defined for both electrons ($A_{m_{e}}$) and holes ($A_{m_{h}}$).}

\begin{tabular}{|c|c|c|c|c|c|}
\hline
alloy Mo$_{2/3}$W$_{1/3}$S$_2$ & \includegraphics[width=2.4cm]{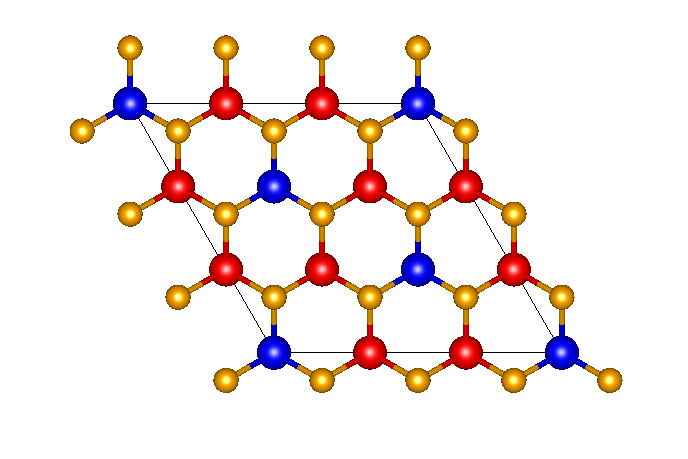} & \includegraphics[width=2.4cm]{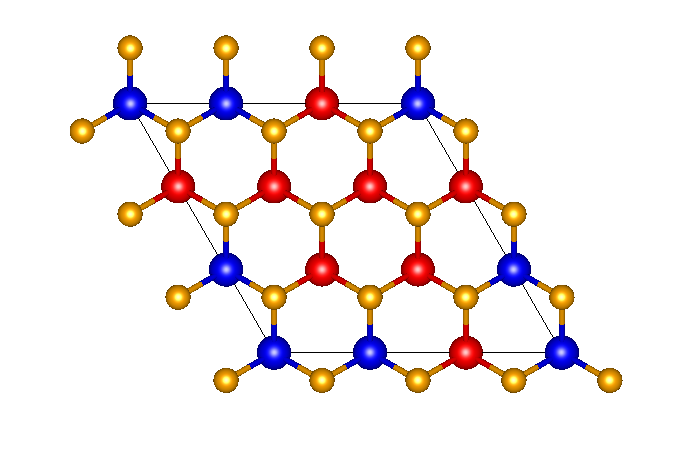} & \includegraphics[width=2.4cm]{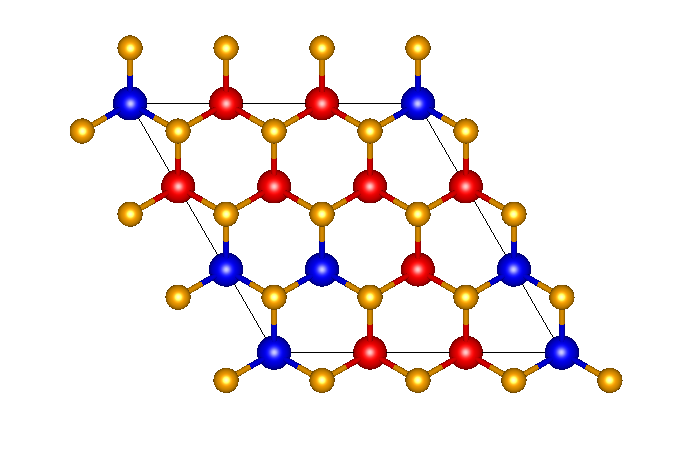} & \includegraphics[width=2.4cm]{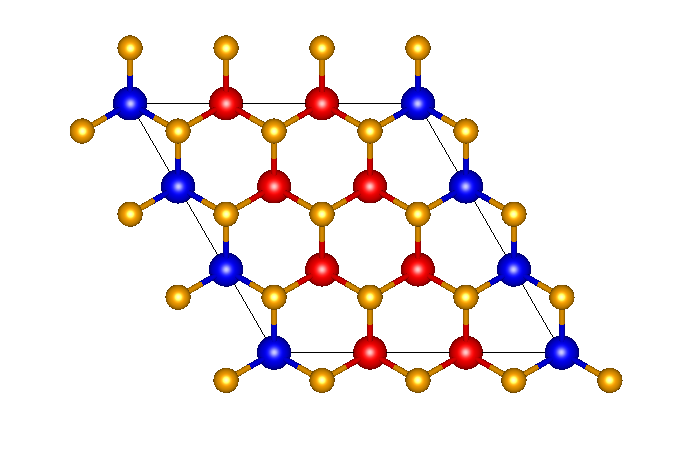} & \includegraphics[width=2.4cm]{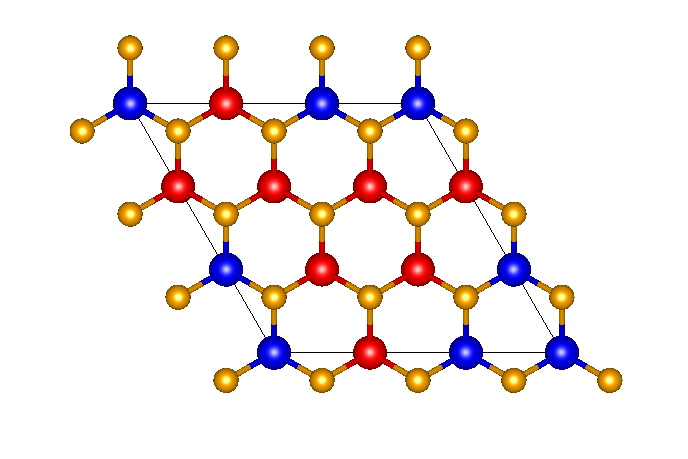} \\ \hline
Configuration & C1 & C2 & C3 & C4 & C5 \\
Space group& P$\overline{6}$2m (189) & Pm (6) & P$\overline{6}$m2 (187) & Amm2 (38) & P$\overline{6}$m2 (187)\\
Primitive cell? &  F & T & T & F & T \\
\hline
$\Delta E^{VBM}_{SP}$/$\Delta E^{VBM}_{SOC}$  [meV] & $<$0.1/235 & 0.7/234 & $<$0.1/234 & $<$0.1/234 & 1.7/234 \\
atom character $w^{VBM}$ & mixed & mixed & mixed & mixed & mixed \\
orbital character VBM& $d_{x^{2} - y^{2}} + d_{xy}$& $d_{x^{2} - y^{2}} + d_{xy}$ & $d_{x^{2} - y^{2}} + d_{xy}$ & $d_{x^{2} - y^{2}} + d_{xy}$ & $d_{x^{2} - y^{2}} + d_{xy}$ \\
$\Delta_{K\Gamma}^{VBM}$ [meV] &243 & 243 & 243 & 243 & 243\\
\hline
$\Delta E^{CBM}_{SP}$/$\Delta E^{CBM}_{SOC}$  [meV] &  267/235 & 115/103 & 0.1/3.4 & $<$0.1/2.6 & $<$0.1/2.6 \\
atom character $w^{CBM}$& Mo-like& Mo-like & mixed & mixed & mixed \\
orbital character CBM & $d_{z^2}$& $d_{z^2}$ & $d_{z^2}$ & $d_{z^2}$ & $d_{z^2}$ \\
$\Delta_{KQ}^{CBM}$ [meV] & 59 & 147 & 202 & 199 & 204\\
\hline
N active transitions & 8 & 8 & 4 & 4 & 4 \\
\hline
$ A_{m_{e}}$ / $ A_{m_{h}}$ & 0.001/0.01 & 0.009/0.01 & 0.005/0.01 & 0.078/0.14 & 0.006/0.01\\
\hline
\end{tabular}
\end{table}

\begin{multicols}{2}

 The more pronounced anisotropy observed for the valence band indicates that hole transport is particularly sensitive to local symmetry breaking, which may also affect recombination processes and optical response. Although such effects would average out in a fully random alloy, their emergence at specific compositions highlights the role of local structural variations. If particular atomic configurations can be stabilized, this anisotropy may be exploited for designing direction-dependent electronic and optoelectronic devices.

\subsubsection*{Configuration-dependent electronic features}
The Table 2.1 summarizes the structural and electronic properties of symmetry-inequivalent configurations at a fixed composition $x = 1/3$. For each configuration (C1--C5), the space group, the possibility of a primitive-cell description, band-edge splittings obtained with and without spin--orbit coupling (SOC), and the atomic and orbital character of the valence-band maximum (VBM) and conduction-band minimum (CBM), the total number of active transitions at $K$ high symmetry point and the anisotropy of effective masses for both types of carriers are reported.

Chemical inhomogeneity partially lifts the degeneracy of the band edges already at the scalar-relativistic level (spin-polarized SP calculations). For the VBM, this splitting remains small ($\Delta E_{\mathrm{VBM}}^{\mathrm{SP}} < 1$~meV), while SOC significantly enhances it to $\sim 234$~meV, with only weak dependence on the atomic configuration. This behaviour is consistent with the mixed Mo and W $d$-orbital character of the VBM, dominated by $d_{x^2-y^2}$ and $d_{xy}$ states. In addition, the calculations were performed using a $3\times3$ supercell, in which the $K$
 point is folded onto the 
$\Gamma$ point. To verify that the folded $\Gamma$-derived states do not mix with the VBM, we evaluated the energy difference between the $K$ and $\Gamma$ points. This separation remains large (243 meV) and nearly constant for all configurations, confirming the absence of significant mixing.

In contrast, the CBM exhibits a pronounced sensitivity to the local atomic arrangement. The scalar-relativistic splittings $\Delta E_{\mathrm{CBM}}^{\mathrm{SP}}$ vary strongly across configurations, ranging from nearly degenerate values ($<0.1$~meV) for C3--C5 up to 267~meV for C1 and C2. Upon inclusion of SOC, the splitting in C3--C5 increases to a few meV, whereas in C1 and C2 it remains at the level of a few hundredths of meV.   In C1, C2 the lowest CB states are dominated by different transition-metal species (Mo-like or W-like), indicating a partially subsystem-like character governed by local symmetry breaking and chemical environment. The comparison between SP and SOC calculations shows that the splitting of the band edges,  is already present at the scalar-relativistic level, indicating a non-relativistic origin associated with chemical inhomogeneity and local symmetry breaking. This is in contrast to pristine transition MoS$_2$ and WS$_2$, where band-edge splitting is predominantly driven by spin–orbit coupling \cite{zhu2011giant, liu2015electronic}.

Additionally, the quantity $\Delta_{KQ}^{\mathrm{CBM}}$ measures the energy separation between the conduction-band minima at the $K$ and $Q$ valleys and thus indicates the stability of the $K$-valley CBM against valley reordering. In pristine MoS$_2$ and WS$_2$, the $K$ and $Q$ valleys are known to be close in energy (see Fig. \ref{Band structures}(c) and  Fig. S1 for the entire composition range in the SI). In the present case, $\Delta_{KQ}^{\mathrm{CBM}}$ remains large (ranging from 59 mev for C1 till 204 meV in C5) and positive for all configurations, showing that the CBM is consistently located at $K$ (with the exception of WS$_{2}$ as noted previously). The substantial variation of this parameter further demonstrates that the relative valley energetics are sensitive to the local atomic arrangement, with C1 lying closest to a possible $K$--$Q$ crossover.
\begin{Figure}
    \centering
\includegraphics[scale=0.10]{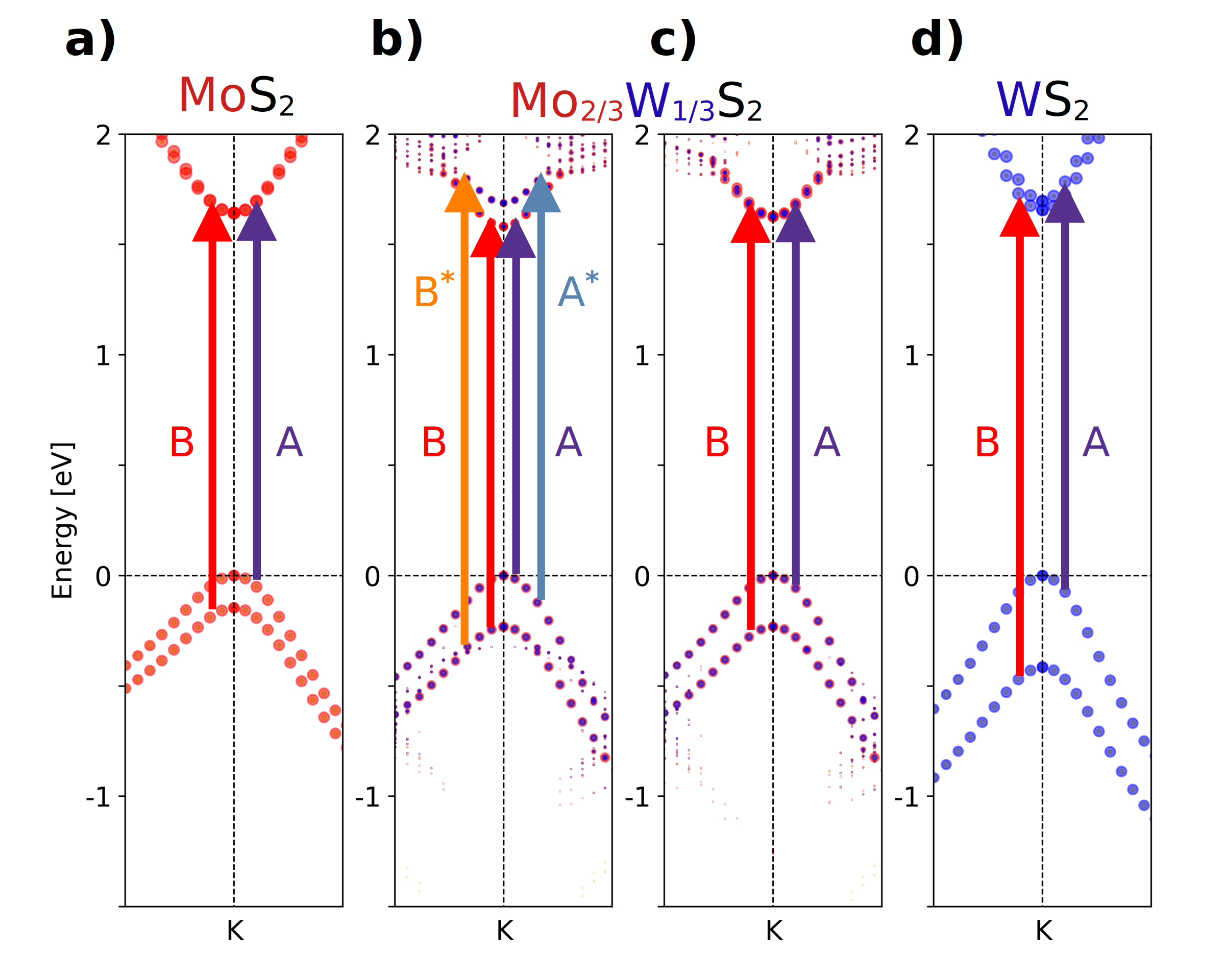}
\captionof{figure}{Schematic illustration of the main optical transitions at the $K$ point of the first Brillouin zone for (a) MoS$_2$, (b) the C2 configuration of Mo$_{2/3}$W$_{1/3}$S$_2$, (c) the C5 configuration of Mo$_{2/3}$W$_{1/3}$S$_2$, and (d) WS$_2$, calculated within PBE+SOC. All transitions are circularly polarized, with $\sigma^{+}$ polarization for the A transition and $\sigma^{-}$ for the B transition, with opposite polarization at the $K'$ point. For selected alloy configurations, additional transitions labelled A$^{*}$ and B$^{*}$ emerge, exhibiting the same polarization as their A and B counterparts. These additional transitions appear when the conduction bands are energetically well separated, which occurs for most configurations across the composition range, except for configurations C3--C5 at $x = 1/3$ and their counterparts at $x = 2/3$.}
\label{transition_scheme}
\end{Figure}

A clear configuration dependence is observed in the number of optically active band-edge transitions (see Table~2.1 and Fig.~\ref{transition_scheme}). Configurations C1 and C2 exhibit eight active VBM $\rightarrow$ CBM transitions (four per valley), whereas only four are found for C3--C5 (two per valley). This behaviour is governed by the splitting of the conduction-band states: well-separated conduction bands lead to eight active transitions, while reduced splitting, observed for selected configurations at $x = 1/3$ and $x = 2/3$, results in only four. Accordingly, configurations with energetically separated conduction bands exhibit additional symmetry-allowed optical transitions, denoted as $A^*$ and $B^*$, alongside the conventional $A$ and $B$ transitions at the $K$ valley. These additional transitions retain the same circular polarization as their parent transitions, reflecting preserved valley-contrasting selection rules. In contrast, configurations with nearly degenerate conduction bands support only the conventional $A$ and $B$ transitions.

Overall, while the VBM remains largely insensitive to local configuration, the CBM and optical properties are strongly influenced by atomic arrangement, highlighting the role of local structural variations in determining the electronic and optical response of the alloy.

\subsubsection*{Absorption spectra}

\begin{Figure}
    \centering
    \includegraphics[scale=0.6]{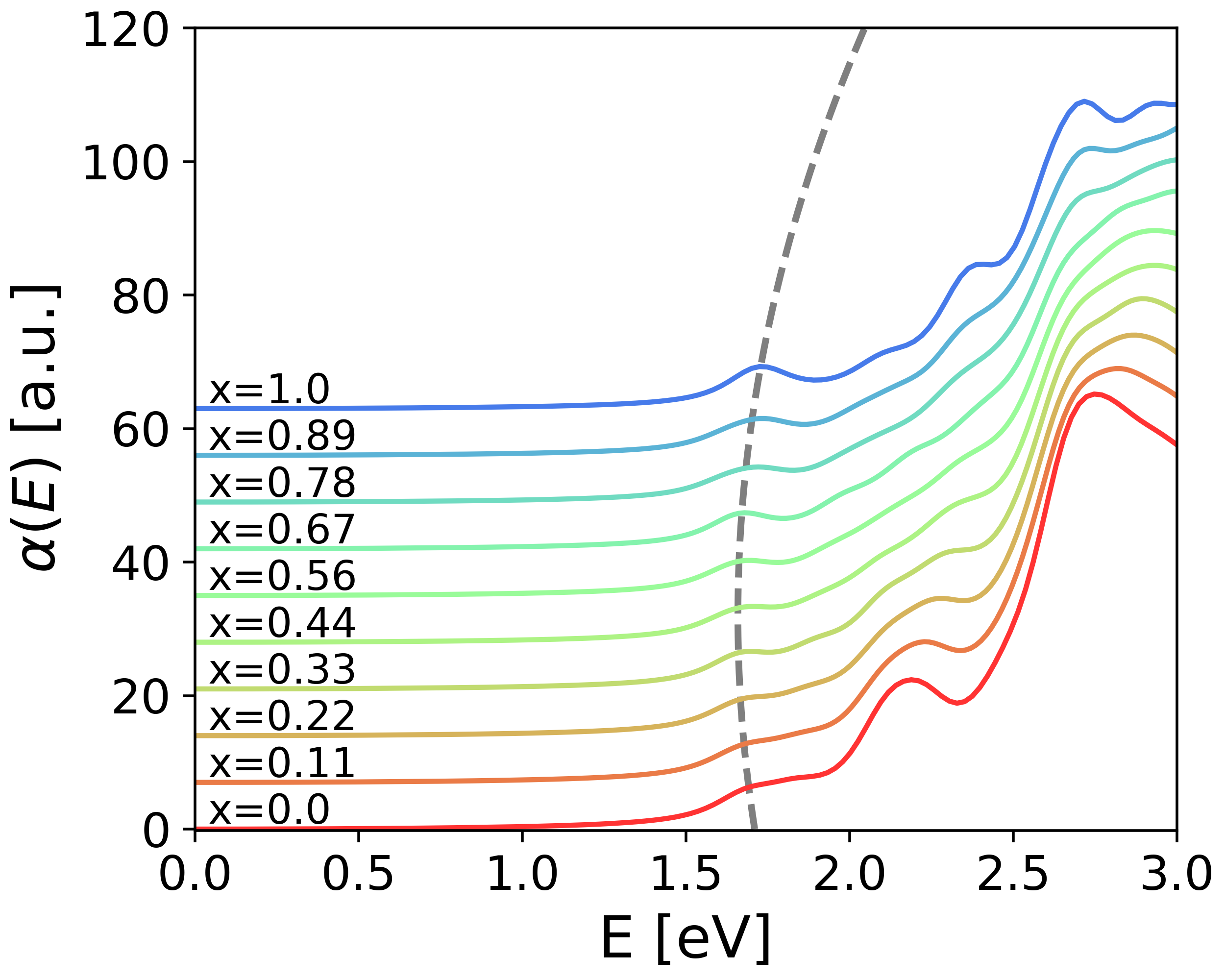}
    \captionof{figure}{Absorption coefficient $\alpha(E)$ calculated from the dielectric function ($\epsilon$) calculated for one representative configuration at each composition. The absorption coefficient remained nearly identical across different configurations, suggesting that this parameter is governed by the global electronic structure rather than local atomic arrangements. While the overall spectral shape remains similar, the characteristic features become progressively broadened and less distinct for intermediate compositions, indicating disorder-induced smearing of the optical transitions. The dashed line indicates the evolution of the absorption onset (band gap) with composition}
    \label{Absorption}
\end{Figure}

The absorption coefficient obtained from the dielectric function shows a systematic evolution with composition. The main absorption onset shifts with $x$, consistent with the composition dependence of the band gap, while the higher-energy peaks become broader and less resolved in the alloyed systems than in pristine MoS$_2$ and WS$_2$. The strongest smearing occurs near the middle of the composition range, where configurational disorder is maximal. This indicates that alloying does not qualitatively change the optical response, but progressively washes out sharp spectral features through disorder-induced broadening.

\section{\normalsize CONCLUSIONS}

We systematically investigated the structural, energetic, electronic, and optical properties of monolayer Mo$_{1-x}$W$_x$S$_2$ alloys using a combination of density-functional theory, an effective excitonic model, and Monte Carlo simulations. The Monte Carlo approach was employed to account for configurational disorder and statistical fluctuations, while DFT provided the electronic structure and band-edge parameters used to construct an effective excitonic Hamiltonian within the Rytova--Keldysh framework.

We find that the cohesive energy varies nearly linearly with composition, indicating that the alloy forms a near-ideal substitutional solid solution, with energetics primarily governed by composition rather than atomic configuration. Configuration-dependent energy differences are small and originate from weak local chemical interactions that slightly disfavour Mo--Mo and W--W nearest-neighbour pairs. Finite-temperature Monte Carlo simulations reveal an order--disorder transition at low temperatures ($\sim 20$~K), above which the alloy is thermodynamically disordered. At experimentally relevant temperatures, Mo$_{1-x}$W$_x$S$_2$ is therefore expected to form a homogeneous alloy with a random distribution of dopants.

The electronic structure exhibits a systematic evolution with composition, with the band gap remaining direct at the $K$ point across the entire range. The band gap follows a non-linear composition dependence characterized by a small bowing parameter, placing the system in the weak-bowing regime typical of chemically similar alloys. We found that in contrast to the weak configurational dependence of the valence-band maximum, the conduction-band minimum shows pronounced configuration-dependent splitting, leading to variations of up to two orders of magnitude and strongly affecting the valley energetics.

These configuration-induced modifications have direct consequences for the optical properties. In particular, the number of optically active band-edge transitions depends sensitively on the local atomic arrangement. Configurations with well-separated conduction bands exhibit additional symmetry-allowed transitions beyond the conventional A and B excitons, whereas nearly degenerate cases show a reduced number of active transitions, observed for specific configurations at $x = 1/3$ and $x = 2/3$. Furthermore, the hole effective masses at the valence-band maximum display configuration-dependent anisotropy, reflecting local symmetry breaking and implying direction-dependent transport properties.

Overall, our results demonstrate that while structural stability and energetics are largely composition-driven, the electronic and optical response is governed by atomic-scale configuration. This highlights the importance of explicitly accounting for configurational effects in TMD alloys and provides guidelines for tailoring their properties for optoelectronic applications.

\section*{Methods}
To investigate ordered Mo$_{1-x}$W$_x$S$_2$ alloys, a $3 \times 3$ supercell of the 2H phase (stable under ambient conditions) was employed, containing 27 atoms. This construction enables 10 distinct compositions $x$, for which various dopant arrangements were considered. Symmetry reduction using the SOD code \cite{SOD} yielded 28 nonequivalent configurations.  

First-principles calculations were performed within density functional theory (DFT) as implemented in the Vienna Ab-initio Simulation Package (VASP) software \cite{VASP1, VASP2}. The Perdew–Burke–Ernzerhof functional was used together with projector augmented-wave pseudopotentials \cite{blochl1994projector, kresse1999ultrasoft}. A kinetic energy cutoff of 400 eV and $\Gamma$-centered $9 \times 9 \times 1$ ($3 \times 3 \times 1$) $k$-point meshes were adopted for pristine (alloyed) systems. Electronic convergence was set to $10^{-8}$ eV.

Lattice parameters for the alloys were obtained via linear interpolation between MoS$_2$ (3.160~\AA) and WS$_2$ (3.153~\AA). Structural optimization was performed by relaxing ionic positions, while spin–orbit coupling (SOC) was included only in subsequent electronic-structure calculations, as its effect on relaxed geometries was negligible but computationally demanding.

Band structures were computed along the high-symmetry path $\Gamma$--M--K--$\Gamma$ and unfolded for supercells using \textit{Vaspkit} software \cite{vaspkit}. Details of the effective-mass and optical-property calculations are provided in sections S1 and S2 of the Supporting Information (SI), respectively. Cluster-expansion and Monte Carlo simulations were performed using \textit{icet}~\cite{icet} (see section S3 of the SI). Exciton binding energies were obtained by solving the Rytova--Keldysh model in real space (section S4 of the SI).

\section*{ Acknowledgments }
We acknowledge financial support from the National Science Centre (NCN), Poland under
the grant no. 2024/53/B/ST3/04258. 
We gratefully acknowledge Polish high-performance computing infrastructure PLGrid (HPC Center: ACK Cyfronet AGH) for providing computer facilities and support within computational grant no. PLG/2025/018673.

\end{multicols}

\printbibliography[
heading=bibintoc,
title={References}
]
\end{refsection}

\clearpage

\begin{center}
    \LARGE \textbf{Supporting Information}
\end{center}

\renewcommand{\thefigure}{S\arabic{figure}}
\setcounter{figure}{0}

\renewcommand{\thesection}{S\arabic{section}}
\setcounter{section}{0}
\renewcommand{\theequation}{S\arabic{equation}}
\setcounter{equation}{0}

\section{ Effective Mass Calculations}

The electronic properties were analyzed by evaluating the effective masses of charge carriers at the K point. The effective-mass tensor $m^{*}$ is defined as
\begin{equation}
\left( \frac{1}{m^{*}} \right)_{ij} = \frac{1}{\hbar^{2}} \frac{\partial^{2} E_n(\mathbf{k})}{\partial k_i \partial k_j}, \quad i,j = x,y,z,
\end{equation}
where $E_n(\mathbf{k})$ is the dispersion of the $n$-th band. The derivatives were evaluated at the band extrema (conduction-band minimum or valence-band maximum), where the effective-mass approximation applies. Second derivatives were obtained numerically using a finite-difference scheme with a five-point stencil and a step of 0.01~Bohr$^{-1}$, as implemented in the \textit{EMC} code~\cite{EMC}. The resulting tensor was diagonalized to obtain the principal components $m^{*}_1$, $m^{*}_2$, and $m^{*}_3$. For the two-dimensional systems considered here, only the in-plane components are relevant for transport, as the out-of-plane dispersion is negligible. This approach captures both isotropic and anisotropic band curvature, enabling a quantitative assessment of direction-dependent carrier dynamics in the alloy systems.

\section{Optical properties}
 Optical transitions at the K point were analyzed by evaluating the momentum matrix elements between valence and conduction states, expressed as $\mathbf{p}_{cv\mathbf{k}}$. The transition probability is proportional to $|\mathbf{p}_{cv\mathbf{k}}|^{2}$, while the polarization is determined by the direction of this vector. In particular, the dipole selection rules for circularly polarized light are governed by $|\hat{e}_{\pm} \cdot \mathbf{p}_{cv\mathbf{k}}|^{2} > 0 \leftrightarrow \sigma^{\pm}$, where $\hat{e}_{\pm} = \frac{1}{\sqrt{2}} (\hat{x} \pm i \hat{y})$ correspond to right- ($\sigma^{+}$) and left-handed ($\sigma^{-}$) circular polarization, respectively. These selection rules originate from the symmetry of pristine monolayer MoS$_2$ and WS$_2$, which crystallize in the $P6m2$ space group, giving rise to valley-selective optical transitions at the K and K$'$ points.

The frequency-dependent optical response was further characterized by the complex dielectric function, $\varepsilon(\omega) = \varepsilon_{1}(\omega) + i \varepsilon_{2}(\omega)$, computed within the independent-particle approximation implmented in VASP, including 400 bands at the PBE+SOC level. Within this approach, electron--hole (excitonic) interactions are not included. The absorption coefficient $\alpha(E)$ was obtained from the dielectric function as
\begin{equation}
\alpha(E) = \frac{\sqrt{2}E}{\hbar c} \left[ \sqrt{\varepsilon_{1}^{2} + \varepsilon_{2}^{2}} - \varepsilon_{1} \right]^{1/2}.
\end{equation}

The calculated absorption spectra were found to be nearly insensitive to the specific atomic configuration at a fixed composition, indicating that local disorder has a minor impact on the overall optical response. Therefore, the spectra presented in Fig.~5 correspond to the lowest-energy configuration for each alloy composition.
\begin{figure}
    \centering
    \includegraphics[scale=0.80]{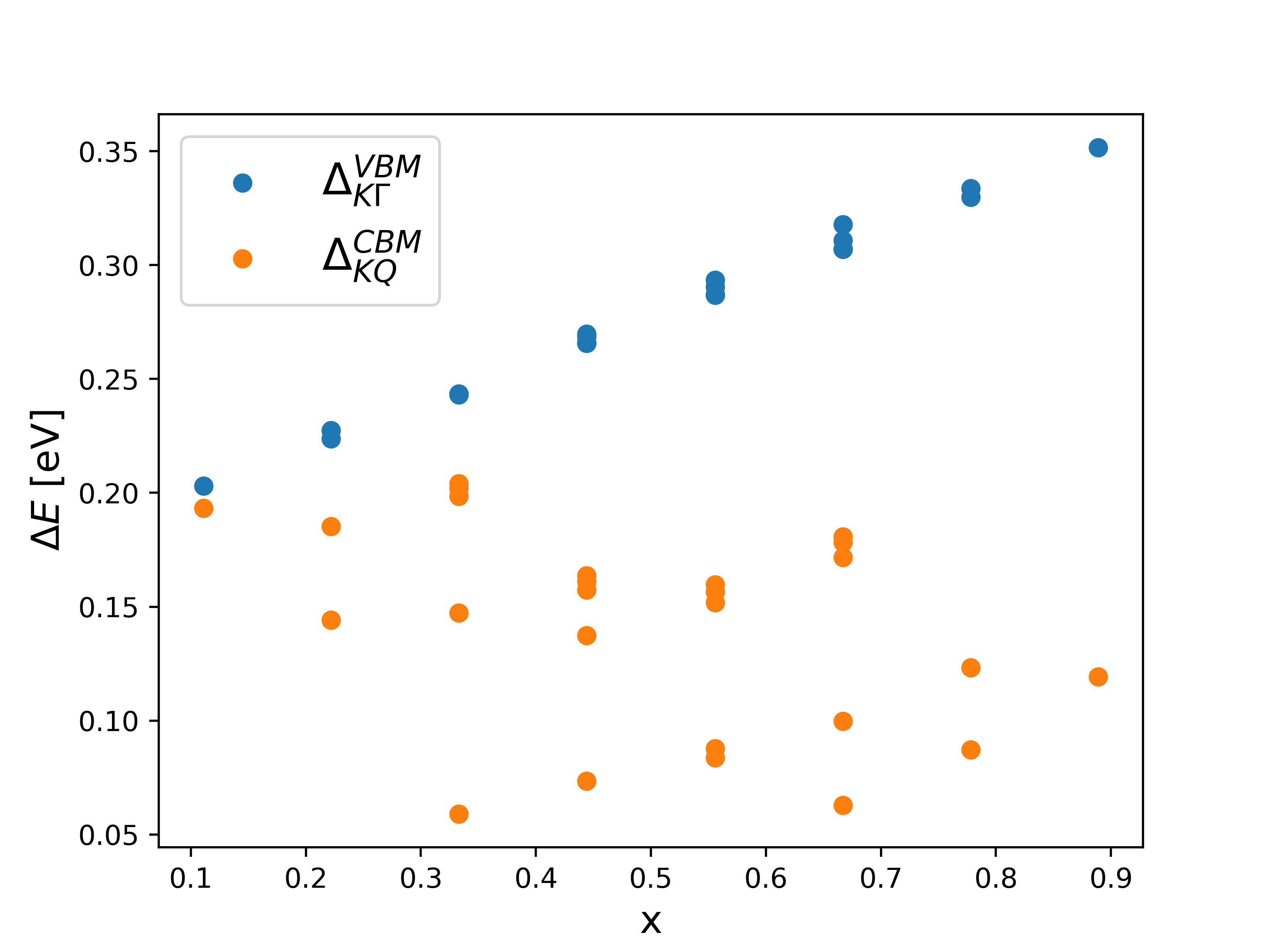}
    \caption{The calculated energy difference between the conduction band points of Q and K (orange), as well as between K and $\Gamma$ points in the valence band (blue). The relation in the valence band follows a linear trend with minor variations between different atomic arrangements, while the conduction band exhibits more complex behaviour, with atomic arrangements dominating the compositional relation.}{}
    \label{KQ_KGamma}
\end{figure}

\section{Cluster Expansion}
To examine the alloy mixing behavior over a wide temperature range, a cluster-expansion (CE) model was constructed using the \textit{icet} software package~\cite{icet}. The model was trained on PBE total energies of all calculated $3\times3$ supercells, supplemented by 215 additional nonequivalent $4\times4$ structures at concentrations $x = 0.25$, 0.5, and 0.75. The inclusion of the larger supercells ensured convergence of the CE parameters.

Only one-, two-, and three-body clusters were included, with cutoff radii of 10~\AA{} for pairs and 4~\AA{} for triplets. Larger clusters yielded negligible contributions, indicating that the interactions are short-ranged. Tests showed that the choice of training method had no significant impact on the results; therefore, Bayesian ridge regression was adopted, with model validation performed using $k$-fold cross-validation ($k=10$).

Monte Carlo simulations were carried out on a $20\times20$ supercell to allow for sufficient configurational sampling. The simulations were initialized with random atomic distributions and performed within the canonical ensemble, keeping the concentration fixed. Temperatures in the range 1--400~K were considered, with $4 \times 10^{4}$ Monte Carlo steps per temperature.

The mixing behavior was quantified using the binary short-range order (SRO) parameter, restricted to nearest neighbors, defined as
\begin{equation}
    \mathrm{SRO} = 1 - \frac{P_{\mathrm{Mo,W}}}{x_{\mathrm{W}}},
\end{equation}
where $P_{\mathrm{Mo,W}}$ is the probability of finding a W atom as a nearest neighbor of a Mo atom, and $x_{\mathrm{W}}$ is the W concentration.

\section{Exciton Calculations within the Rytova–Keldysh Model}

Exciton binding energies were evaluated within the Rytova-Keldysh framework by solving the effective two-dimensional electron--hole Hamiltonian in real space. The interaction between charge carriers was described by the screened potential
\begin{equation}
    V(r) = -\frac{e^2}{8\epsilon_{0}r_{0}}\left[H_{0}\left(\frac{r}{r_{0}}\right) - Y_{0}\left(\frac{r}{r_{0}}\right)\right],
\end{equation}
where $e$ is the elementary charge, $\epsilon_{0}$ is the vacuum permittivity, and $r_{0} = 2\pi \chi_{2D}/\epsilon_{\mathrm{env}}$ is the screening length, with $\chi_{2D}$ being the two-dimensional polarizability obtained from the dielectric response of the monolayer. $H_{0}$ and $Y_{0}$ denote the Struve and Bessel functions of the second kind (order zero), respectively.

The kinetic-energy operator was discretized using a three-point finite-difference scheme. Tests with higher-order stencils yielded identical eigenvalues within numerical accuracy, while increasing computational cost.

The resulting radial Hamiltonian was solved on a uniform real-space grid spanning $10^{-13}\,\mathrm{m} < r < 5r_{0}$ with a step size of $5 \times 10^{-4} r_{0}$. Convergence with respect to grid spacing and domain size was explicitly verified. The discretized Hamiltonian assumes a tridiagonal form and was diagonalized to obtain excitonic eigenvalues.

\begin{figure}
    \centering
    \includegraphics[scale=0.12]{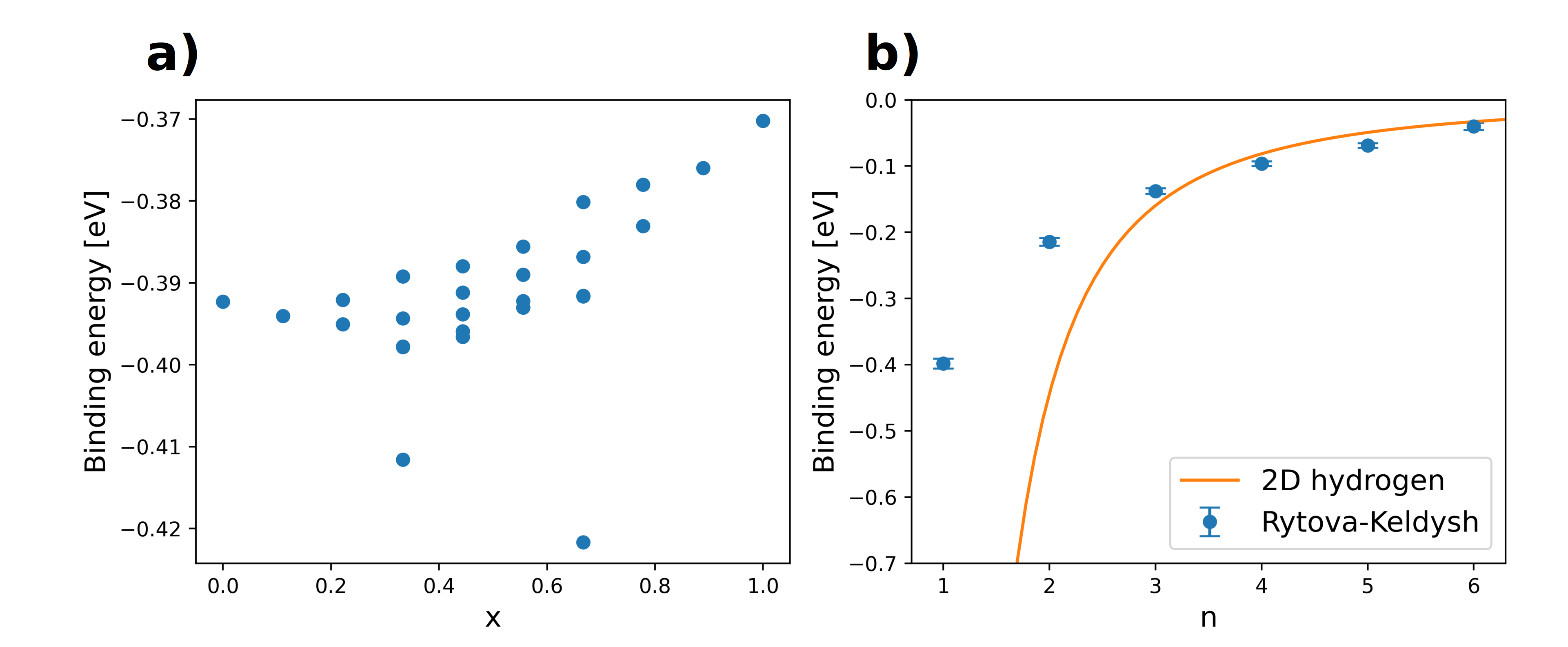}
    \caption{The results of the Rytova-Keldysh model for Mo$_{(1-x)}$W$_{x}$S$_{2}$. (a) The dependence of the exciton binding energy depending on the composition and atomic arrangements. The binding energies follow a non-linear dependence with large deviations for anisotropic configurations. Slight differences can be observed between different atomic configurations. (b) The mean energy of the states of the exciton for x=1/3 as a function of quantum number $n$. The energy of the first two states deviates significantly from the 2D hydrogen model solution (screened 2D Coulomb potential) $E(n) = -E_{b}/(n-1/2)^2$}{}
    \label{excitons1}
\end{figure}

The calculated binding energies, shown in Fig.~\ref{excitons1}(a), exhibit a nonlinear dependence on composition, consistent with the bowing observed in the band gap. Variations between configurations at fixed concentration arise primarily from differences in the carrier effective masses, while the screening (via $\chi_{2D}$) remains nearly composition dependent but configuration independent. Pronounced deviations are observed for highly anisotropic configurations, reflecting their larger effective masses. It should be noted that the Rytova--Keldysh model assumes an isotropic reduced mass $\mu$ and therefore does not explicitly capture mass anisotropy.

The lowest excitonic states are presented in Fig.~\ref{excitons1}(b). Their energy spectrum deviates significantly from the idealized two-dimensional hydrogenic model, $E(n) = -E_b/(n - 1/2)^2$, particularly for the lowest states ($n=1,2$), highlighting the importance of nonlocal dielectric screening in two-dimensional systems.

\newpage

\printbibliography[
heading=bibintoc,
title={References}
]

\end{document}